\documentclass[preprintnumbers,aps,prl,twocolumn,superscriptaddress,showpacs,amsmath,amssymb]{revtex4}

\usepackage{graphicx}% Include figure files
\usepackage{dcolumn}% Align table columns on decimal point
\usepackage{bm}% bold math

\newcommand{\undemi}{\frac 1 2 }

\newcommand{\elas}{^{\mathrm{e}}}
\newcommand{\trans}{^{\mathrm{t}}}

\newcommand{\IN}{_{\mathrm{IN}}}
\newcommand{\NW}{_{\mathrm{NW}}}
\newcommand{\IW}{_{\mathrm{IW}}}

\newcommand{\nvec}{\mathbf{n}}
\newcommand{\lvec}{\mathbf{l}}
\newcommand{\mvec}{\mathbf{m}}

\newcommand{\Qvec}{\mathbf{Q}}
\newcommand{\Qij}{Q_{i,j}}
\newcommand{\Acal}{\mathcal{A}}
\newcommand{\Scal}{\mathcal{S}}

\newcommand{\dd}{\mathrm{d}}
\newcommand{\ed}{\mathrm{e}}

\DeclareMathOperator{\Tr}{Tr}

\begin{document}

\preprint{CTP-PP-JMRE/v4.3}

\title{Complete wetting transitions of nematic liquid crystals on a structured substrate }% Force line breaks with \\

\author{Chi-Tuong Pham}
\email[]{pham@cii.fc.ul.pt}
\affiliation{Centro de F{\'\i}sica Te\'orica e Computacional, 
Universidade de Lisboa,
Avenida Professor Gama Pinto 2, P-1649-003 Lisboa Codex, Portugal}
%\homepage{http://www.Second.institution.edu/~Charlie.Author}
\author{Pedro Patr\'icio}
\affiliation{Instituto Superior de Engenharia de Lisboa
Rua Conselheiro Em\'\i dio Navarro 1, P-1949-014 Lisboa, Portugal}
\affiliation{Centro de F{\'\i}sica Te\'orica e Computacional, 
Universidade de Lisboa,
Avenida Professor Gama Pinto 2, P-1649-003 Lisboa Codex, Portugal}
\author{Jose Manuel Romero-Enrique}
\affiliation{Departamento de F\'\i sica At\'omica, Molecular y Nuclear, Area de F\'\i sica Te\'orica
Universidad de Sevilla, 
Apartado de Correos 1065, 41080 Sevilla, Spain
}%

\date{\today}% It is always \today, today,
             %  but any date may be explicitly specified

\begin{abstract}
In this article, we generalize Wenzel law, which assigns an effective contact angle for a droplet on a rough substrate,
when the wetting layer has an ordered phase, like a nematic.
We estimate the conditions for which the wetting behavior of an ordered fluid can be qualitatively 
different from the one usually found in a simple fluid.
To particularize our general considerations, we will use the Landau-de Gennes mean field approach to 
investigate theoretically and numerically the complete wetting transition between a nematic liquid crystal
and a saw-shaped structured substrate.

\end{abstract}

% PACS, the Physics and Astronomy
                             % Classification Scheme.
%\keywords{Suggested keywords}%Use showkeys class option if keyword
                              %display desired
\maketitle

It is known that the wetting behavior of a fluid is deeply altered in
the presence of rough or structured substrates.  Leaves for instance have often
developed a patterned texture, with micro reliefs, in order to adapt
themselves to a particular humid environment \cite{Neinhuis_Barthlott}.
Recently, technological advances allowed the
controlled manufacturing of artificial micro structured substrates,
which were used to show spectacular results concerning water-repellency, 
switchable wettability, and other practical applications \cite{Callies_Quere_2005}.\par
In this article, we will investigate the wetting behavior of an ordered fluid (a nematic liquid crystal),
on a rough substrate.
We will first review some simple considerations about isotropic fluids and rough substrates,
and then we will generalize these ideas for the case of ordered fluids.
We will particularize our study by considering the complete wetting of a nematic on a saw-shaped substrate.
Quantitative results will be obtained by solving analytically and numerically the 
Landau-de Gennes free energy.\par
\begin{figure}[t] 
\centerline{\includegraphics[width=.3\textwidth]{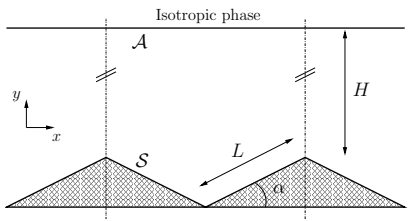}}
\caption 
{Geometry of the system. $\Acal$ is the projected surface of the wedged (rough) surface $\Scal$}  
\label{fig2_geometry}
\end{figure}

Consider two isotropic phases at coexistence (let us call them $A$ and $B$
phases), their bulk free energy densities being the same $f_A=f_B=0$.
Suppose now the $B$-phase is the one preferred in the far field, and our system is in the presence of a flat substrate or wall.
The substrate interacts with the fluid through a local surface energy with strength $w$, which favors the $A$-phase.
In this situation, an $A$-phase layer may appear close to the wall, because the decrease we have in the surface energy is already
sufficient to compensate the creation of an interface between the two phases. Let us define the wettability function $g(w)$ as 
%\begin{equation}
$g(w)=\sigma_{BW}-\sigma_{AW}$ 
%\label{eq1}
%\end{equation}
where $\sigma_{\alpha \beta}$ is the surface tension associated to a \emph{flat} $\alpha$-$\beta$ interface.
For fixed bulk coexistence conditions, the wettability coefficient will depend on the strength of the surface energy, and usually is an
increasing function of $w$. The Young equation yields  $g(w)=\sigma_{AB}\cos \theta_\pi$ where  $\theta_\pi$ is the contact angle of the sessile drop.
Thus, as $w$ increases, the contact angle $\theta_\pi$ decreases. Eventually, $\theta_\pi=0$ at the wetting transition, when the $A$-$B$ interface
unbinds from the substrate. In this case $g(w=w_\pi\trans)=\sigma_{AB}$, where $w_\pi\trans$ is the transition value. For larger values of $w$, the interface 
remains unbounded as a thick $A$-phase layer is formed between the substrate and the bulk $B$-phase (complete wetting).
The specific effective interactions between the wall and the substrate determine the order of the wetting transition as well as the
role played by interfacial capillary wave fluctuations \cite{Sullivan_1986,Dietrich_1988,Forgacs_1991}.
Wetting on rough substrates presents a richer phenomenology than for flat substrates. Interfacial unbinding may occur via a sequence of different phase 
transitions like unbending (or filling) and unbinding \cite{Rascon_1999,Rejmer_2000}. However, a simpler picture arises if we assume wetting as a one-step transition. Then one can easily predict the wetting behavior of this fluid on a 
mesoscopically rough substrate by thermodynamic arguments. The partially filled phase has an excess free energy which can be related to the surface 
tension $\sigma_{BW}$
as $\Delta F_1=\Scal\sigma_{BW}$, where $\Scal$ is the substrate surface area. At the complete filled situation, the excess free energy is given by $\Delta F_2
= \Scal\sigma_{AW} + \Acal \sigma_{AB}$, where $\Acal$ is the $A$-$B$ interface area, which coincides with the surface area of projection of the substrate onto the
tangent plane (see Fig. \ref{fig2_geometry}). Wetting transition occurs for $\Delta F_1=\Delta F_2$, or equivalently for $g(w=w_r\trans) = \sigma_{AB}/r$, where 
$w_r\trans$ is the surface coupling at the wetting transition for the rough substrate and $r\equiv \Scal/\Acal>1$ is the substrate roughness. This result is 
consistent with Wenzel law \cite{Wenzel_1936}, which assigns an effective
contact angle for the sessile droplet on the rough substrate $\theta_r$ as:
%\begin{equation}
$\cos \theta_r =r \cos \theta_\pi.$
%\label{Wenzel}
%\end{equation} 
So, the wetting transition occurs for $\theta_r\to 0$, in agreement with our previous result. Since $g(w)$ is an increasing function of $w$, we find
generally that $w_r\trans<w_\pi\trans$. \par
This picture can be changed dramatically when the wetting layer has an
ordered phase, as in the case of a nematic liquid crystal. For this case,
the rough or structured substrate may impose a deformation on the
ordered fluid, which must be accompanied by a positive elastic free energy
$\Delta F\elas_{AW}>0$. Note that the elastic deformation creates an effective
long-range repulsion between the substrate and the wall: the closer to the
substrate the interface is, the more constrained the order will be, leading to 
higher energies. Consequently, first-order wetting will be more likely to occur in this case.  
Again the wetting transition may be obtained by the
free energy balance $\Acal\sigma_{AB}+\Scal\sigma_{AW}+\Delta F\elas_{AW}=\Scal\sigma_{BW}$.
Note that $\sigma_{AW}$ is the equilibrium surface tension associated to 
the ordered fluid-flat substrate in absence of any order parameter deformation.   
The critical surface field for a rough substrate, $w_r\trans$, has to verify the relation
$g(w_r\trans)=\sigma_{AB}/r + \Delta F\elas_{AW}/\Scal$.  For the wetting of ordered fluids,
$w_r\trans<w_\pi\trans$ if distortions are not very important; $w_r\trans>w_\pi\trans$ if the
energy of the distortions is sufficiently large.  If the effects of
the elastic deformations are too strong, the energy balance may never
be favored, for all values of the surface field $w$. In this last
case, there will be no wetting transition. 

%An important point should be made here, in what regards the nature of
%the wetting transition for ordered fluids.  The elastic deformation
%creates an effective long range interaction between the substrate and
%the interface. If the interface is closer to the substrate, the order
%will be necessarily more constrained, leading to higher energies.  At
%phase coexistence, we predict then a first order wetting transition,
%between a non wetting regime and an infinite thick layer. This
%argument will be used below to simplify our analysis.

We will particularize these considerations by choosing a periodic
saw-shaped substrate characterized by an angle $\alpha$ (see Fig. \ref{fig2_geometry}).
The plane of the paper is defined by the axes $x$ and $y$. Out of the
plane, the wedge is parallel to the $z$-axis. 
%For simple fluids, thermodynamic arguments
%are not enough to locate the wetting transition in this geometry \cite{Rejmer_2000}. Mean-field analysis
%of the interfacial effective Hamiltonian \cite{Rejmer_2000,Rejmer_2004} shows that for the first-order wetting case,
%the wetting transition occurs at a temperature between the corresponding transition values for a planar substrate
%and the predicted by Wenzel law. For continuous wetting there is no shift of the wetting transition. However, 
%a filling transition emerges at the temperature predicted by Wenzel law (i.e. $\cos \theta_\pi =\cos \alpha$) for 
%$L\to \infty$ \cite{Rejmer_2000,Concus_1969,Pomeau_1986,Hauge_1992}. 
In our geometry, we
defined two lengths, the length of the wedge side, $L$, and the height
between the substrate and the isotropic ($B\equiv I$) phase at the far-field,
$H$.  Close to the substrate, a new nematic ($A\equiv N$) phase may or not
appear. In fact, for the complete wetting transition we are
discussing, the length $H$ will be unimportant, because the
wetting layer, if it exists, has an infinite thickness.

Both isotropic and nematic phases can be represented by the Landau-de
Gennes tensor order-parameter $\Qij$.  Owing to the traceless and
symmetric character of the tensor order-parameter, it can be very
generally represented by 
%\begin{equation}
$\Qij = \frac 3 2 S[n_in_j - \frac 1 3 \delta_{i,j}] + \frac 1 2 B
[l_il_j - m_i m_j]$
%\label{tensor}
%\end{equation}
where $S$ is the scalar order parameter ($S=0$ in the isotropic phase
and $S\neq 0$ when some order is present), and $B$ is the biaxiality~\cite{Andrienko_Tasinkevytch_Patricio_etal_2004}.
The direction of maximal orientational order is given by the director
$\nvec$, and the unit vectors $\nvec,\lvec,\mvec$ form a local
orthonormal triad.

In our problem, we will only consider in-plane deformations, although
out-of-plane or twist deformations may also be important (a twist
instability may occur for particular choices of parameters \cite{Patricio_Telo_Dietrich_2002}).  
In this case, $\nvec=(\cos\theta,\sin\theta,0)$,
and the tensor order parameter has three independent components only,
$Q_{11}$, $Q_{22}$, and $Q_{12}$.

The system is described by the Landau-de Gennes free energy
%\begin{equation}
$\mathcal{F}_{\mathrm{LdG}} = \int_{\mathcal{V}} (f_{\mathrm{bulk}} +
f_{\mathrm{el}})\,\dd V + \int_{\mathcal{S}} f_{\mathrm{surf}}\,\dd s$
%\label{free_energy}
%\end{equation}
where $f_{\mathrm{bulk}}$ is the bulk free energy density,
$f_{\mathrm{el}}$ is the elastic free energy density~\cite{Pgg}, and
$f_{\mathrm{surf}}$ is the surface free energy.  Here, we will take
the commonly used rescaled expressions \cite{Andrienko_Tasinkevytch_Patricio_etal_2004}:
\begin{align}
& f_{\mathrm{bulk}} = \frac 2 3 \tau \Tr \Qvec^2 - \frac 8 3 \Tr
\Qvec^3 + \frac 4 9 [\Tr \Qvec^2]^2&&\\ & f_{\mathrm{el}} = \frac {\xi^2}
%1 
{3+2\kappa}[\partial_k \Qij\partial_k \Qij + \kappa \partial_j \Qij
\partial_k Q_{i,k}] &&\\ & f_{\mathrm{surf}} = - \frac 2 3 w \Tr
[\Qvec\cdot\Qvec_{\mathrm{surf}}]
\end{align}
where $\tau$ is a dimensionless temperature, $\kappa$ is an elastic
dimensionless parameter, $w$ is a dimensionless surface field or
anchoring strength, and $\xi$ is the so-called correlation length that
will be set to $1$ in the rest of the article.  For $\tau=1$, the two
phases are at coexistence, $f_{\mathrm{bulk}}$ has two minima, which
correspond to the scalar order parameters $S_\mathrm{I}=0$ (isotropic
phase) and $S_\mathrm{N}=1$ (nematic phase).  The elastic parameter is
restricted to the values, $\kappa>-3/2$.  If $\kappa$ is positive
(negative), the nematic prefers to align parallel (perpendicular) to
the nematic-isotropic interface.  Also, $\Qvec_{\mathrm{surf}}$
defines the favored tensor at the substrate.  In our problem, we will
favor a homeotropic (or perpendicular) alignment of the nematic
at the substrate, and a bulk nematic scalar order
parameter ($S_{\mathrm{surf}}=S_\mathrm{N}=1$).  This particular
choice for the surface free energy was made in order to establish a
direct connection to previous related papers \cite{Sheng_1976,Braun_1996}.  
%We note that there is only one relevant physical
%parameter in the bulk energy density, because two rescalings were
%made, redefining the total free energy and the tensor order
%parameter. Also, there is only one physical parameter in the elastic
%energy density, because we rescaled the coordinate space.

Consider first the flat substrate ($\alpha=0$), already numerically
studied in the literature \cite{Sheng_1976,Braun_1996}.  At
coexistence, $\tau=1$, it is straightforward to estimate each term of
the balance equation.  The non-wet isotropic configuration can be
simply represented by the scalar order parameter profile
$S(y)=a\ed^{-y/y_0}$, and $\theta=\pi/2$.  
%In these analytical
%calculations, we will neglect the eventually small existing biaxiality (when $\theta=0$).
Introducing this function in the total free energy $\mathcal{F}_{\mathrm{LdG}}$, we obtain
%\begin{equation}
$\mathcal{F}_{\mathrm{LdG}}/\Acal= \int_0^\infty (S^2-2S^3+S^4+\undemi
S'^2)\dd y-wS(0)$
%\label{energy_IW_NW}
%\end{equation}
where $\Acal$ is the area of the substrate, and $S'=\dd S/\dd y$.  The free
energy is minimized when $y_0=1/\sqrt{2}+O(w)$, and
$a=w/\sqrt{2}+O(w^2)$.  If we use this solution, we obtain a very
accurate value for the surface tension between the substrate and the
isotropic phase in the non-wet configuration 
$\sigma\IW =-w^2/(2\sqrt{2})+O(w^3)$.  In order to
calculate the surface tension between the substrate and the nematic
phase, we may assume that the nematic director is everywhere oriented 
perpendicular to the substrate ($\theta=\pi/2$), and that the scalar order
parameter profile is described by $S(y)=a\ed^{-y/y_0}+S_\mathrm{N}$. 
Following the same steps as before, we find $\sigma\NW =-w-w^2/(2\sqrt{2})
+O(w^3)$. Finally, the nematic-isotropic surface tension can be estimated 
through the ansatz $S(y)=\undemi [1-\tanh(y/2y_0)]$ and assuming either the 
director is everywhere oriented either perpendicular ($\theta=0$) or parallel 
($\theta=\pi/2$) to the interface, which are the relevant situations
for $\kappa<0$ and $\kappa>0$, respectively. If we introduce this
function in the total free energy, and perform the integration over
the whole real $y$-axis, we obtain the energy of the interface,
depending on one free parameter $y_0$. Minimization will give
$y_0=1/\sqrt{2}$, and the surface tension between the isotropic and
the nematic phase is $\sigma\IN^\perp =\sqrt{2}/6\approx
0.236$ ($\theta=0$) and $\sigma\IN^\parallel =\sigma\IN^\perp
\sqrt{\frac{3+\kappa/2}{3+2\kappa}}$ ($\theta=\pi/2$).
In the latter calculation, we neglected the existing biaxiality. 
The balance equation may in this case be written as
\begin{equation}
g(w)=\sigma\IW-\sigma\NW=w+O(w^3)=\sigma\IN
\label{bal_eq_analytic}
\end{equation}
where $\sigma\IN=\sigma\IN^\perp$ for $\kappa<0$ and $\sigma\IN=
\sigma\IN^\parallel$ for $\kappa>0$. In the latter, the elastic contribution
due to the director distortions in the nematic film due to the mismatch of the
anchoring conditions at the substrate and the nematic-isotropic interface
should be included in the
balance Eq. (\ref{bal_eq_analytic}). However, this contribution vanishes 
with the nematic film thickness $H$ as  
$\Delta F\elas/\Acal \sim 1/H$, so it can be safely ignored at the wetting
transition, where $H\to \infty$. 
If for example $\kappa=2$, the surface field for the
complete wetting transition is $w_B=0.18\pm 0.01$ (see
Eq. (\ref{bal_eq_analytic})), which is in excellent agreement with the 
numerical value obtained by Braun \emph{et al.} \cite{Braun_1996}.
Note that if we take $\sigma\IN=\sigma\IN^\perp$, the transition surface
field is $w_S=0.24\pm 0.01$, which is the numerical value obtained by
Sheng \cite{Sheng_1976}.

We now turn to the rough substrate.
Using Landau-de Gennes free energy,
it is easy to see that the elastic contribution scales as $\Delta F\elas=K\ell\Delta \tilde F\elas$,
where $K$ is the Frank elastic energy and $\ell$ is the typical length of the rough substrate. 
If the nematic-isotropic interface goes to infinity, $\Delta\tilde F\elas$ is a number only dependent on
the substrate geometry, but not on its scale $\ell$. To calculate the critical surface field for the wetting transition,
we have the generalization of Wenzel law:
\begin{equation}
g(w_r\trans)=\frac{1}{r}(\sigma\IN + \frac{K}{\ell}\Delta \tilde F\elas%_{NW}
)
\label{Generalized_Wenzel_Law}
\end{equation}
This equation is the main result of this work. For large enough substrate lengths $\ell$,
the effects of the elastic terms are not important, and Wenzel law is recovered.
However, this law is significantly changed when $\ell\sim K\Delta\tilde F\elas/\sigma\IN$.
In the Landau-de Gennes model, $K$ and $\sigma\IN$ are not independent. For $\tau=1$ (phase coexistence),
$K /(\xi \sigma\IN)\approx 2.6/0.2 $.
The elastic numerical factor was calculated for a sinusoidal grating \cite{Berreman_1972,Patricio_Telo_Dietrich_2002},
$\Delta \tilde F\elas=(2\pi)^3/4 a^2$,
where $a$ is the ratio between the amplitude and the wavelength, $l$ of the sinusoid.
For our wedge geometry, we may use $\tan \alpha=4a$, and $\alpha=\pi/4$,
to estimate $L\sim 35 \xi$ (see Fig.~2 for definition of $L$), for which the elastic effects still play an important role.

To calculate the critical surface field, $w\trans$, for every angle $\alpha$ of our geometry,
we numerically minimized the Landau-de Gennes free energy, using
a conjugate-gradient method. The numerical discretization of the continuum problem was performed
with a finite element method combined with adaptive meshing in order to
resolve the different length scales \cite{Patricio_Tasinkevych_Telo_2002}.
As before, we restricted ourselves to the case of coexistence ($\tau=1$) and set $\kappa=2$.
We have imposed periodic boundary conditions (b.c.) on $x$ in every 
calculation and assumed translational invariance along the $z$ direction,
so the problem reduces to an effective two-dimensional case.

To obtain the surface energy $\sigma\IN^{\perp / \parallel}$, isotropic ($S=0$) fixed b.c.
were imposed at the top of the domain and homeotropic ($\perp$) or parallel ($\parallel$) nematic ($S=1$) fixed b.c. at the bottom.
The energy $F\IW$ of the non-wet isotropic configuration 
was calculated by imposing isotropic fixed b.c. only at the top of the domain.
At the bottom, there were no imposed b.c., and the surface free energy was taken into account.
Finally, for the calculation of the energy $F\NW$ of a bulk nematic phase in
contact with the substrate, fixed nematic b.c. were imposed at the top,
with the director angle either set to $\theta=\pi/2$ (perpendicular case)
or to $\theta=0$ (parallel case), and different cell heights $H$ were 
considered.

\begin{figure}[t]
\centerline{\includegraphics[angle=-90, width=.4\textwidth]{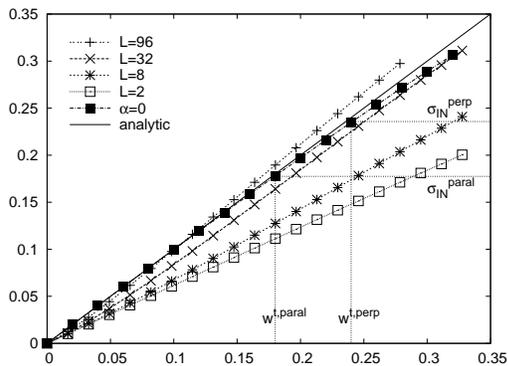}}
\caption{Plot of $(F\IW(\alpha, L, w)-F\NW(\alpha, L, w))/(2L\cos\alpha)$
vs. $w$ at angle $\alpha=35^\circ$ together with the flat case $\alpha=0$ 
and analytical results (Eq. (\ref{bal_eq_analytic})). The slope
increases with increasing $L$ so that the transition surface field 
diminishes. %Inset: Dependence of the slope at origin with respect to
%angle $\alpha$ for $L=32$. The slope decreases for increasing $\alpha$
%up to $\alpha = 90^\circ$ so that the critical anchoring increases for
%increasing $\alpha$. 
}
\label{fig_slopes}
\end{figure}

As an important check for our procedure, we recovered 
Braun's and Sheng's values of the transition surface field for $\alpha=0$:
$w_\pi^{\mathrm{t},\parallel}\simeq 0.1796 $ and $w_\pi^{\mathrm{t},\perp}\simeq 0.2417$, 
by solving numerically the equation $\sigma\IN^{\perp / \parallel}=
\sigma\IW -\sigma\IN$. 
The agreement with our theoretical results is excellent as the $w^3$ corrections  
in $g(w)$ can be neglected (see Fig. \ref{fig_slopes}).

We now turn to the case where $\alpha$ is non zero. 
As expected for the non-wet isotropic configuration, we obtain that $F\IW(\alpha, L, w)\sim\sigma\IW(w)\times 2L$, 
where $\sigma\IW(w)$ is the substrate-isotropic surface tension in the flat case. This result is essentially independent of the cell height $H$. 
On the other hand, $F\NW$ depends crucially on the geometrical parameters of the substrate,
$\alpha$ and $L$. Numerical minimization shows that the global minimum configuration  
corresponds to the solution for $\theta=0$ at the top cell boundary for $\alpha\le\pi/4$ and the minimum energy solution 
for $\theta=\pi/2$ at the top cell boundary for $\alpha\ge\pi/4$. In both branches the energy $F\NW$ decreases with the cell height $H$, and converge to the relevant free energy 
to the complete wetting transition.
These solutions mean that for $\alpha<\pi/4$ the nematic wetting film configuration is an hybrid aligned 
nematic (HAN), where the nematic director is oriented along the $y$ direction above the rough substrate, and changes smoothly to the planar
anchoring at the NI interface. For $\alpha>\pi/4$, the nematic film is essentially parallel (P) to the NI interface above the substrate. 
As the HAN and the P configurations are metastable for $\alpha>\pi/4$ and $\alpha<\pi/4$, respectively, a first-order HAN-P 
transition occur at $\alpha=\pi/4$, analogous to that observed for sinusoidal substrates \cite{Harnau_2006_2}.   

%For given angle $\alpha$ and $L$, the plots of the energy difference
%$(F\IW(L, w)-F\NW(\alpha, L, w))/(2L\cos\alpha)$ do not differ
%qualitatively from those found in the flat case except that the
%dependence on $L$ at a given angle $\alpha$ is such that the slope
%increases with increasing $L$. %Also, the slopes of the curves
%decrease with increasing $\alpha$ up to $\pi/4$% %(see inset of Fig. \ref{fig_slopes}) for a given $L$
%. %Note that this latter figure is the result of the measurement of the slope at origin of the curves. 

\begin{figure}[t]
\centerline{\includegraphics[angle=-90,width=.4\textwidth]{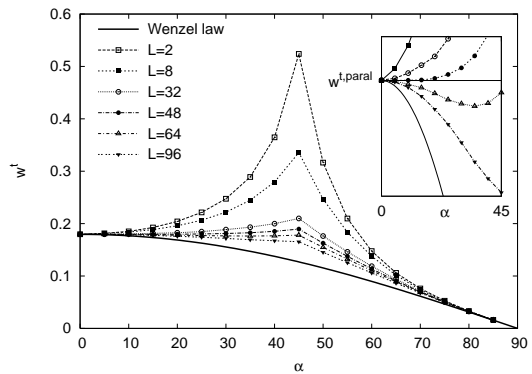}}
\caption{Transition surface field  $w_r\trans$
for different values of $L$  together with Wenzel law extrapolated from the flat case. For $L\gtrsim 48$,  the convexity of the curve changes:  roughness effects can lower $w_r\trans$ compared to the flat case. For larger $L$, Wenzel law is expected to be recovered. Inset : blowup of the curve.}
\label{fig_results_alpha}
\end{figure}

In order to compute the transition surface field at which
complete wetting occurs, we solve the equation
$(F\IW-F\NW)/(2L\cos\alpha)=\sigma\IN^{\parallel}$ (equivalent
to Eq. (\ref{Generalized_Wenzel_Law})) for different values of $\alpha$ and 
$L$. The left-hand side function does not differ
qualitatively from those found in the flat case except that the
dependence on $L$ at a given angle $\alpha$ is such that the slope
increases with increasing $L$ (see Fig. ~\ref{fig_slopes}). Note that the smaller the slope, the higher the
transition surface field. The transition surface fields are displayed in Fig.
\ref{fig_results_alpha}. The kink observed at $\alpha=\pi/4$ is a consequence of the HAN-P transition at the nematic wetting film.
For moderate values of $L$ ($L<48$), we can see that the transition surface field for a given $L$ increases with
$\alpha$ up to $\alpha=\pi/4$, where the transition value $w_r\trans$ is larger than the corresponding one for the flat case, 
and decreases for larger $\alpha$. 
For $L\gtrsim 48$, a change in the convexity of the curve at the origin occurs (see inset of  Fig.  \ref{fig_results_alpha}): a new regime 
is reached where roughness favors complete wetting. For larger values of $L$ (e. g. $L=96$) the critical anchoring always 
decreases with $\alpha$, and one should expect the elastic energy contributions to vanish for $L\to \infty$ and Wenzel law $w_r\trans(\alpha)=
w_\pi^{\mathrm{t},\parallel} \cos\alpha$ to be recovered. Our numerical results confirm that prediction. They also show that for arbitrary $L$ the 
deviation of the transition surface field with respect to that predicted by Wenzel law is larger around $\alpha=\pi/4$, implying that the
nematic director field is always more deformed for $\alpha=\pi/4$ than for any other angle.   

In this article, we generalized Wenzel law for nematics.  We used
the Landau-de Gennes model to investigate theoretically and
numerically the complete wetting transition between a nematic liquid
crystal and a saw-shaped structured substrate.  %In future
%works, we intend to generalize our results for higher temperatures.
One should keep in mind that the typical length of the structured
surface should be only of order 30 $\xi$ in order to observe large deviations from the Wenzel law.  
At these scales, it is not
clear whether the mean field approach we are assuming still
holds. Other physical phenomena may be present in real
situations. Confirming these results experimentally may be a
challenge.

\begin{acknowledgments}
\textbf{Acknowledgements: }P. P. thanks T. Sluckin and M. M. Telo da Gama for enlightening discussions. C.-T. P. acknowledges the support of Funda\c c\~ao para a Ci\^encia e
Tecnologia (FCT) through Grant No. SFRH/BPD/20325/2004. J.M. R.-E. acknowledges a ``Ram\'on y Cajal'' Fellowship and financial support from Junta de Andaluc\'{\i}a (Ayudas PAIDI FQM-205).

\end{acknowledgments}
%\nocite{*}
\bibliography{wedge}

\end{document}